\documentstyle[amstex,amssymb,preprint,aps]{revtex}

\begin{document}
\title{Phenomenological model for symmetry breaking in chaotic system}
\author{A. Y. Abul-Magd and M. H. Simbel}
\address{Faculty of Science, Zagazig University, Zagazig, Egypt}
\date{\today}
\maketitle

\begin{abstract}
We assume that the energy spectrum of a chaotic\ system undergoing symmetry
breaking transitions can be represented as a superposition of independent
level sequences, one increasing on the expense of the others. The relation
between the fractional level densities of the sequences and the symmetry
breaking interaction is deduced by comparing the asymptotic expression of
the level-number variance with the corresponding expression obtained using
the perturbation theory. This relation is supported by a comparison with
previous numerical calculations. The predictions of the model for the
nearest-neighbor-spacing distribution and the spectral rigidity are in
agreement with the results of an acoustic resonance experiment.
\end{abstract}

\section{INTRODUCTION}

Random matrix theory provides a framework for describing the statistical
properties of spectra for quantum systems whose classical counterpart is
chaotic \cite{mehta,guhr1}. It models the Hamiltonian of a chaotic system by
an ensemble of $N$-dimensional random matrices, subject to some general
symmetry constraints. Time-reversal-invariant quantum system having integral
spins are represented by a Gaussian orthogonal ensemble (GOE) of random
matrices while those having half-integral spins are modeled by a Gaussian
symplectic ensemble (GSE). Chaotic systems without time reversal invariance
are represented by the Gaussian unitary ensemble (GUE). Symmetries
associated with quantum numbers always involve a structure of the
Hamiltonian matrices, which is reflected in the composition of the energy
spectra. When the system has such a symmetry with $M$ eigenvalues, the
Hamiltonian of the system is block diagonal. Each block represents an
eigenvalue (or a set of eigenvalues) of the symmetry operator, and may be
considered as a member of one of the above mentioned canonical ensembles.
The energy spectrum is given by a superposition $M$ of independent
sequences, each one representing one of the Hamiltonian blocks.
Modifications accounting for the symmetry breaking are realized by
introducing coupling blocks belonging to different quantum numbers. The
first attempt in this direction is due to Rosenzweig and Porter \cite
{rosenzweig}. Their model was expanded by French et al \cite{french} who
applied the perturbation theory to study transitions between different
classes of symmetry. Guhr and Weidenm\"{u}ller \cite{guhr} applied this
model to study mixing of states with isospin $T=0,1$ in nuclei. Leitner et
al. \cite{leitner,leitner1} obtained closed-form expressions for the
nearest-neighbor spacing (NNS) distributions within its framework by
applying the perturbation theory.\ Their results has been applied to
experimental spectra for two coupled microwave billiards \cite{alt} by
Barbosa and Harney \cite{barbosa}. One however may have difficulties with
this model when the number of the symmetry eigenvalues $M$ is more than two.
More information has to be put in the theory since the influence of symmetry
breaking on different eigenfunctions may be different. Moreover, one may
have difficulties in applying Leitner's perturbation formulae for symmetries
with large $M$, which are valid for energy intervals of length considerably
exceeding $M$. Numerical investigations for breaking such symmetries require
diagonalizing larger matrices, which are built of sizable blocks, in order
to achieve a statistical significance comparable to that of modern
experiments. Additionally, the number of allowable symmetry eigenvalues is
not always fixed, as in the case of multiple collective excitation of
nuclei. For example, the maximum number of phonons realized in the nuclear
vibrational model is subject to dispute \cite{chomaz}.

In the present paper we consider a simple model for gradual symmetry
breaking in a chaotic system. This model represents the energy spectrum as
independent level sequences not only in the absence of the symmetry breaking
perturbation, but also during the whole transition until the symmetry is
totally violated. During the crossover from full to no symmetry, one of the
sequence grows on the expense of the other sequences until it totally
exhausts the spectrum. The proposed model leads to approximate expressions
for the nearest-neighbor and next-to-nearest-neighbor spacing distributions 
\cite{dembo} which depends on a single parameter, namely the mean level
density of the sequences. These have been successfully applied to the
analysis of the spectra of coupled microwave resonators \cite{alt}, which
are not described by a Hamiltonian divided into a symmetry breaking term and
a perturbation. The model provides a satisfactory description for level
statistics of low-lying 2$^{+}$ states of even-even nuclei \cite{2plus}
without explicit knowledge of their symmetry properties. This paper presents
additional arguments in favor of that model. Section II shows that the
energy spectrum obtained in the Rosenzweig-Porter model \cite{rosenzweig}
may approximately be represented as a composite spectrum of level sequences.
Section III uses the asymptotic values for the expressions of level-number
variance obtained by French et al \cite{french} for systems undergoing a
symmetry breaking to find a relation between the fractional level density of
the sequences and the symmetry-breaking perturbation strength. This relation
is tested in Section IV by a comparison of the NNS distributions of the
composite spectrum with the ones obtained by Leitner \cite{leitner} through
a diagonalization of the Rosenzweig-Porter Hamiltonian. Section V shows that
the model under investigation is consistent with the results of the
acoustic-resonance experiment by Andersen et al. \cite{andersen}. The
conclusion and summary of this work are given in Section VI.

\section{A MODEL FOR SYMMETRY BREAKING}

\ The purpose of this section is to show that the level spectrum of a
chaotic system with a conserved symmetry can still be expressed as a
superposition of independent sequences, even when the symmetry is partially
violated . For concreteness, we restrict the consideration in this section
to the transition from two independent GOE spectra to a single GOE spectrum.
The extension to other universality classes is trivial. We assume that the
system is described by the Rosenzweig-Porter model \cite{rosenzweig}. The
Hamiltonian of the perturbed system can be written as a sum of a
block-diagonal matrix, representing the case when the symmetry is conserved,
and a perturbation responsible for symmetry breaking. When the symmetry has
two eigenvalues of same degeneracy, the Hamiltonian takes the form 
\begin{equation}
H=\left( 
\begin{array}{cc}
H_{1} & 0 \\ 
0 & H_{2}
\end{array}
\right) +\varepsilon \left( 
\begin{array}{cc}
0 & V \\ 
V^{\dagger } & 0
\end{array}
\right) ,
\end{equation}
where $H_{1}$ and $H_{2}$ are GOE matrices with elements having a variance $%
v^{2}(1+\delta _{ij})$, and $V$ is a random matrix with elements having the
same variance so that $\varepsilon =1$ makes $H$ as a whole to be a GOE
matrix.

Because the system is chaotic, the eigenvalues of the symmetry are expected
to spread out over phase space. Thus, all of the matrix elements that couple
different eigenstates are statistically equivalent. Keeping this in mind, we
consider a simple version of the Rosenzweig-Porter model in which $%
H_{1}=H_{2}$. hoping that the obtained results approximately apply for
independent diagonal Hamiltonian blocks. We start by the case in which the
symmetry is fully conserved ($\varepsilon =0$). The total wavefunction is
given by a product of an eigenfunction of the symmetry operator and an
eigenfunction that depends on all coordinates except the one that represents
the symmetry. For concreteness, we shall refer to these as the spatial
coordinates, although other variables may be involved. In the case under
consideration, the symmetry operator $S$ has two eigenvalues, say $s_{1}$
and $s_{2}$. It can be represented by the $2\times 2$\ matrix, diag$%
(s_{1},s_{2}),$ with corresponding eigenvectors $\alpha _{1}$ and $\alpha
_{2}$. \ Wavefunctions of a chaotic system have nearly a uniform phase-space
distribution. Nothing practically happens to the spatial components of the
wave functions by introducing the symmetry breaking perturbation. \ In this
case, the effect of symmetry breaking can approximately be taken into
account by simply introducing a new symmetry operator $S_{\delta }$ that
contains additional non-diagonal elements, \ each equal to $\delta $, which
is proportional to $\varepsilon $. The wavefunction of each perturbed state
is given by a spatial wavefunction that does not depend on the symmetry
breaking, multiplied by one of the two eigenfunctions $\beta _{1,2}$ of $%
S_{\delta }$ having eigenvalues $\lambda _{1,2}=\frac{1}{2}\left[
s_{1}+s_{2}\pm \sqrt{(s_{1}-s_{2})^{2}+4\delta ^{2}}\right] .$ Thus, while
the perturbed Hamiltonian (1) has non-vanishing off-diagonal blocks in the
representation in which the symmetry wavefunctions are $\alpha _{1,2}$, it
will recover the block-diagonal form when the representation that involves
the eigenvectors $\beta _{1,2}$\ of $S_{\delta }$ 
\begin{equation}
H=\left( 
\begin{array}{cc}
H_{1}^{^{\prime }} & 0 \\ 
0 & H_{2}^{^{\prime }}
\end{array}
\right) _{\delta }.
\end{equation}
Here, the diagonal blocks $H_{1,2}^{^{\prime }}$ are nearly equal to the
diagonal blocks $H_{1,2}$\ of the unperturbed Hamiltonian, except for a
shift in the diagonal matrix elements by almost the same amount in each
block, but in opposite directions. This shift will produce a corresponding
shift in the eigenvalues corresponding to each block. If the level density
of each sequence in the absence of perturbation is $\rho (E)$, then after
perturbation, the level densities of the sequences become $\rho (E\pm
a\varepsilon )$, respectively, where $a$ is a constant. Now, if $\rho (E)$
is approximated by an exponential function as in many applications, then the
fractional density of the sequence that decreases by the influence of
perturbation, say sequence labeled by 1, can be approximately related to the
perturbation parameter $\varepsilon $ by 
\begin{equation}
f_{1}=f_{1,0}e^{-A\varepsilon },
\end{equation}
where $f_{1,0}$\ is the initial fractional density and $A$ is a constant.

The spectrum of the chaotic system under consideration can approximately be
represented by a superposition of two independent sequences all the way of
the symmetry-breaking transition, even in the case when $H_{1}\neq H_{2}$\
in Eq. (1). Analysis of the spacing distribution of coupled resonators of
unequal size has shown this \cite{dembo}.\ The purpose of the following
sections is to provide further support for this approximation.

We note that the proposed representation of symmetry breaking is meant for
chaotic systems. It does not apply, for example, to the experiment by
Ellegaard {\it et al.}~\cite{ellegaard}, where a gradual breaking of a
point--group symmetry in monocrystalline quartz blocks is achieved by
cutting small pieces at one of the angles. The properties of quartz allowed
these authors to consider the pure sequences as pseudointegrable
systems(see, e.g., Refs.~\cite{richens,biswas,date}). A previous analysis 
\cite{alaa} of this experiment have shown that the symmetry breaking can be
explained by introducing a new GOE level sequence in addition to the
original pseudointegrable ones, and allowing it to grow on their expense.

\section{CALCULATION OF LEVEL-NUMBER VARIANCE}

The level number variance for a spectrum composed of a superposition of $M$
independent sequences is given by \cite{brody} 
\begin{equation}
\Sigma _{\beta }^{2}(r,f_{1},...,f_{M})=\sum_{m=1}^{M}\Sigma _{\beta
}^{2}(f_{m}r),
\end{equation}
where $\beta =1,2$ or $4$ depending on whether the pure sequences belong to
GOE, GUE or GSE, respectively, $f_{m}$ is the average fractional density of
the $m$th sequence, and $\Sigma _{\beta }^{2}(r)$ is the number variance for
the pure sequence. A similar expression is proposed by Seligman and
Verbaarschot \cite{seligman} for the spectral rigidity $\Delta _{3}$, 
\begin{equation}
\Delta _{3\beta }(r,f_{1},...,f_{M})=\sum_{m=1}^{M}\Delta _{3\beta }(f_{m}r)
\end{equation}
French et al. \cite{french} considered the gradual symmetry breaking by the
influence of a perturbation represented by a random matrix with elements $%
H_{ij}$. They calculated the two-level correlation function using the
perturbation theory. They expressed the level number variance as 
\begin{equation}
\Sigma _{\beta }^{2}(r,\Lambda )=\Sigma _{\beta }^{2}(r)+\frac{M-1}{\beta
\pi ^{2}}\ln \left[ 1+\frac{\pi ^{2}r^{2}}{4\left( \tau _{\beta }+\pi
^{2}\Lambda \right) ^{2}}\right] .
\end{equation}
where $\tau _{\beta }$ is a cutoff parameter and $\Lambda $ is the mean
square value of the perturbation measured in units of the mean level spacing 
$D$, 
\begin{equation}
\Lambda =\left. \overline{H_{ij}^{2}}\right/ D^{2}=\varepsilon
^{2}v^{2}/D^{2}.
\end{equation}
The cutoff parameter is estimated by the requirement that, when $\Lambda =0$%
, $\Sigma _{\beta }^{2}(r,0)=\Sigma _{\beta }^{2}(r,f_{1},...,f_{M})$ as
given by Eq. (4). Imposing this requirement means the $\tau _{\beta }$
depends on the energy interval $r$. In previous application of this
approach, e.g. in \cite{leitner,barbosa,dembo}, an average over the range of 
$r$ covered by the data is taken. We here suggest, instead, to determine the
cutoff parameter by requiring the equality of Eqs. (4) and (6) at large
value of $r$, such that $r\gg 1/\min (f_{m})$. For this purpose, we use the
asymptotic expression for the number variance, which is given by 
\begin{equation}
\Sigma _{\beta }^{2}(r)\backsim \frac{2}{\beta \pi ^{2}}\ln \left( 2\pi
r+c_{\beta }+\gamma +1\right) ,
\end{equation}
where $\gamma \cong 0.5772$ is Euler's constant and $c_{\beta }=-\pi
^{2}/8,0,\ln 2+\pi ^{2}/8$ for GOE, GUE, GSE, respectively. Combining Eqs.
(4),\ (6) and (8), we obtain 
\begin{equation}
\tau _{\beta }=C_{M}e^{-(c_{\beta }+\gamma +1)}.
\end{equation}
where $C_{M}=\frac{1}{4}\left( \prod_{m=1}^{M}f_{m}\right) ^{\frac{1}{1-M}}$%
. The pre-exponential factor becomes $C_{M}=\frac{1}{4}M^{\frac{1}{1-1/M}}$\
in the case when all the initial sequences have the same level density $%
f_{m}=1/M$. In particular, if $M=2$, then $\tau _{\beta }=0.709,0.207$ and
0.082 for GOE, GUE and GSE, respectively. In the case of 2 and 8 GOE
sequences of equal initial level density, Leitner \cite{leitner} finds $\tau
_{1}=0.70$ and 1.85 which are in good agreement with our result since Eq.
(9) yield $\tau _{1}=0.709$ and 1.609 for $M=2$ and $8.$

We now assume that the symmetry breaking transition proceeds in such a way
that one sequence, say with $m=I,$ grows on the expense of the others. Then, 
$\Sigma _{\beta }^{2}(r,\Lambda )$ will again be expressed as a sum of
contributions of the $m$ sequences 
\begin{equation}
\Sigma _{\beta }^{2}(r,\Lambda )=\sum_{m=1}^{M}\Sigma _{\beta }^{2}\left[
f_{m}(\Lambda )r\right] ,
\end{equation}
where $f_{m}(\Lambda )$ is the average fractional density of the $m$th
sequences in the presence of perturbation. We further assume, for a
perturbation strength measured by $\Lambda $, the fractional level density
of the sequences with $m\neq I$ decay with the same rate, so that 
\begin{equation}
f_{m}(\Lambda )=f_{m}e^{-\xi _{\beta }(\Lambda )},
\end{equation}
where $\xi (\Lambda )$ is a monotonously increasing function of $\Lambda $.
The condition that $\sum_{m=1}^{M}f_{m}(\Lambda )=1$ yields 
\begin{equation}
f_{I}(\Lambda )=1-(1-f_{I})e^{-\xi _{\beta }(\Lambda )}.
\end{equation}

We can estimate the function $\xi _{\beta }(\Lambda )$\ again by comparing \
Eqs. (4) and (6) in the asymptotic region of large $r$. In the case when all
the initial sequences have equal level densities, $f_{m}=1/M$, this
comparison yields 
\begin{equation}
\xi _{\beta }\left( \Lambda \right) =\pi \sqrt{\frac{2\Lambda }{M\tau
_{\beta }}},
\end{equation}
which agrees with the estimate in Eq. (3), when $A=\pi v/D\sqrt{\tau _{\beta
}}$. This finding shows the consistency of the argument leading to the model
under consideration with the results of the perturbation theory concerning
symmetry breaking. It also allows expressing the fractional level densities
during the symmetry-breaking transition, $f_{m}(\Lambda )$, to the strength
of the symmetry breaking interaction (7).

\section{CALCULATION OF NNS DISTRIBUTION\newline
}

The NNS distribution of a spectrum resulting from a random superposition of $%
M$ independent sequences is well known \cite{mehta}, as mentioned above. Let 
$p_{j}(s)$ denote the NNS distribution for the $j$th subsequence . We now
assume that each of the distributions $p_{j}(s)$ is that of a GOE. The
generalization to other symmetry classes is straightforward. To an excellent
approximation, the $p_{j}(s)$'s are then given by the Wigner surmise. We
define the associated gap functions $E_{j}\left( s\right) =\int_{s}^{\infty
}ds^{\prime }\int_{s^{\prime }}^{\infty }p_{j}(x)dx$ for the subsequences
and the gap function $E\left( s\right) =\int_{s}^{\infty }ds^{\prime
}\int_{s^{\prime }}^{\infty }p(x)dx$ for sequence $S$. Mehta~\cite{mehta}
has shown that $E\left( s\right) =\prod_{j=1}^{M}E_{j}\left( f_{j}s\right) $%
. The NNS distribution for the composite spectrum is obtained by
differentiating $E(s)$ twice. When the symmetry has two eigenvalues, the
spectrum is that of a superposition of two independent GOE level sequences
of fractional densities $f_{1}$ and $f_{2}=1-f_{1}$. The corresponding NNS
distribution is given by 
\begin{multline}
P(s,f_{1})=\frac{\pi }{2}f_{1}^{3}se^{-\pi f_{1}^{2}s^{2}/4}\text{erfc}\left[
\frac{\sqrt{\pi }}{2}\left( 1-f_{1}\right) s\right] \\
+\frac{\pi }{2}\left( 1-f_{1}\right) ^{3}se^{-\pi (1-f_{1})^{2}s^{2}/4}\left[
\text{erfc}\left( \frac{\sqrt{\pi }}{2}f_{1}s\right) \right]
^{2}+2f_{1}(1-f_{1})e^{-\pi \left( 2f_{1}^{2}-2f_{1}+1\right) s^{2}/4}.
\end{multline}
The expression for the spacing distribution becomes more complicated when $%
M>2$. We can obtain an approximate expression for the NNS distribution,
valid for arbitrary $M$, that depends on a single parameter. This is the
mean fractional level density $f=\sum_{i=1}^{M}f_{i}^{2}$ for the
superimposed subsequence, where $f_{j}$\ are the fractional densities of the
constituting ones. \ Some steps in this direction have already been
previously taken \cite{mer3,mer4,dembo,2plus}.

The model proposed above, which represents the spectrum of a system with
partial symmetry as a superposition of independent sequences, suggests
applying Eq.(14) for the transition from the two-GOE statistics to that of a
single GOE. The fractional density $f_{1}$ of the decaying sequence will
then play the role of a tuning parameter. A weak point of the distribution
in Eq. (14) is that it differs from zero at $s=0$, because the
symmetry--breaking interaction lifts the degeneracies. The model thus fails
in the domain of small spacings as far as the NNS distributions are
concerned. The magnitude of this domain depends on the ratio of the strength
of the symmetry--breaking interaction to the mean level spacing. However,
this defect does not affect the long-range statistics (e.g., $\Sigma ^{2}$
or $\Delta _{3}$).\ 

Equations (11) and (13) provide a relation between the fractional density of
the decaying sequence and the symmetry breaking strength. We now test this
relation. Leitner \cite{leitner} numerically diagonalized sets of
real-symmetric matrices of the form in Eq. (1), where the matrices $H_{1}$, $%
H_{2}$ are independent GOE matrices of equal dimension and $V$ contains
interrelated Gaussian random variables. Each set corresponds to a different
value of strength parameter $\varepsilon $ (or $\Lambda $, given by Eq. (7))
and consists of about 2000 matrices of dimension 400. The NNS distribution
obtained by Leitner for four values of $\Lambda =4.05\times
10^{-4},8.11\times 10^{-3},4.05\times 10^{-2},8.11\times 10^{-1},$ are shown
as histograms in Figure 1. The figure compares these
statistically-significant numerical results with the prediction of Eq. (14),
with $f_{1}$ calculated by inserting these values of $\Lambda $ into
Eqs.(11) and (13). We see that the proposed model presents a satisfactory
agreement with the numerical results while using the same values for the
parameter $\Lambda $. \ The only disagreement is between the calculated and
measured values of $P(s)$ at small values of $s$, as we have already
expected.\ The symmetry breaking decreases the probability of finding
degenerate levels sharply, leading to the observed dip at small $s$ in the
spacing distributions of the numerical experiment.\ This dip is followed by
an overshoot to restore normalization. \ The width of this dip, which is a
measure of level-splitting responsible for degeneracy removal, increases
with increasing the parameter $\Lambda $ as expected. In spite of this, we
shall find in the next section the the parameters obtained in the comparison
of the NNS distribution (14) with experiment can successfully be used in the
analysis of other statistics like rigidity $\Delta _{3}$ for the same
spectra.

\section{COMPARISON WITH EXPERIMENT}

Andersen et al. \cite{andersen} measured the frequency spectrum and the
widths for acoustic resonances in thin aluminum plates, cut in the shape of
a three-leaf clover with outer and inner radii 80 mm and 70 mm,
respectively. This shape is chosen because a similar billiard has a chaotic
classical dynamics \cite{jarzynski}. Due to the mirror symmetry through the
middle plane of the plate, each resonance of the plate belongs to one of two
mode classes. The flexural modes, which have displacement mainly normal to
the plane of the plate, are antisymmetric with respect to reflection through
the middle of the plane. The in-plane modes are symmetric. The authors
separated the modes according to their measured widths and showed that each
mode class obeyed the GOE statistics. The number of observed modes in each
class was nearly the same. They introduced a gradual breaking of the mirror
symmetry by cutting a slit of increasing depth on one face of the plate.
They were able to describe the transition that takes place as the mode
classes mixed in terms of a random matrix model.

This section demonstrates that the resonances for the three-leaf clover
plate can approximately be described as two uncoupled classes even if the
symmetry is partially broken. Figure 2 shows by histograms the experimental
width distribution for the resonances when the plate is intact, and when
three different symmetry-breaking splits are cut out. The first case in
naturally described as an independent superposition of contributions from
the in-plane and flexural modes. The authors of \cite{andersen} assume that
the width distribution in this case is given by a linear superposition of
two Gaussian distributions of same weight, i.e. 
\begin{equation}
P(\Gamma )=Cg(\Gamma _{I}^{0},\sigma _{I}^{0})+(1-C)g(\Gamma _{F}^{0},\sigma
_{F}^{0}),
\end{equation}
where $g(\Gamma ,\sigma )$ is a Gaussian distribution with mean value $%
\Gamma $ and variance $\sigma ^{2}$, and the suffixes $I$ and $F$ are for
the in-plane and flexural modes, respectively. The statistical weights of
both classes are set equal, i.e. $C=1/2$.\ They analyzed the width spectrum
of the distorted plates using a random matrix model that describes the
coupling of the two mode classes. Figure 2 analyzes the width distribution
for both the intact and distorted plates by the sum of two Gaussians in Eq.
(15). The statistical weights, mean widths and variances for each mode are
determined by a $\chi ^{2}$-fit. The figure shows that, the overlap of the
width distributions of the two classes indeed increases with symmetry
violation. However, the resolution of the two modes by means of Eq. (15) is
possible even in the case where the symmetry is almost completely destroyed.
Interestingly, the statistical weights of the two modes in all cases remains
the same, equal to 1/2, within the statistical error.

Figure 3 compares the resonance spacing distributions measured in \cite
{andersen} with the prediction of Eq. (14). The fractional density of the
decreasing sequence $f_{1}$ is set equal to 0.5 in the case of impact plate,
where the symmetry is conserved, and considered as a fitting parameter in
the cases of deformed plates. The best-fit values are 0.221$\pm 0.008$, 0.085%
$\pm 0.004$ and 0.047$\pm 0.002$. The agreement with the experiment is good.
An exception is the domain of small spacings (the first bin), where the
model predicts non-vanishing distribution at $s=0$.

Figure 4 compares the experimental values of the $\Delta _{3}$ statistic for
the same four cases considered in Fig. 3 with the ones calculated for a
corresponding superposition of two sequences. The fractional densities of
the sequences used in the calculations are the best-fit values obtained for
the NNS distributions. The agreement between the theoretical curves and
experimental histograms is good. This good agreement means that the
evaluation of the fractional densities of the sequences constituting the
spectrum using Eq. () for the NNS distribution is accurate notwithstanding
the wrong behavior of this distribution at small spacings.

\section{CONCLUSION}

We propose a model for the spectral fluctuations of systems with partially
conserved symmetry. When the symmetry is exact, the spectrum is composed of
a superposition of independent level sequences, each corresponding to a
fixed value of the symmetry quantum number. We argue that the same
representation may still be valid when the symmetry is violated. The
symmetry-breaking transition is modeled by assuming that one of the
sequences is growing on the expense of the others. A relation between the
fractional level densities of the intermediate sequences and the symmetry
breaking interaction strength is obtained by comparing the asymptotic
behavior of the level number variance for the sequence superposition with
the previous results obtained by applying the perturbation theory. The model
is tested by comparing its prediction with the results of numerical
diagonalization of a Hamiltonian divided into a symmetry conserving term and
a perturbation and the outcome of an acoustic resonance experiment. It is
found to give accurate representation for the spectra except in the domain
of small spacing where the symmetry-breaking interaction removes all
possible accidental degeneracies.

The proposed model is not meant to replace more sophisticated models that
describe the breaking of known symmetries, such as isospin or parity.
However, \ it is useful in cases when an approximate symmetry is unknown or
ignored. It has recently been successfully applied to study the NNS
distribution of low-lying $2^{+}$ states of even-even nuclei. Relatively
small values for the mean fractional level densities of the superimposed
sequences are obtained for nuclei expected to have one of the dynamical
symmetries of the interacting Boson model, indicating that their spectra may
be divided into two or more nearly independent sequences.

Table I. Parameters used in the comparison of the width distribution with a
superpesiton of two Gaussian functions, shown in Fig. 2.

\begin{tabular}{|l|l|l|l|l|}
\hline
& Intact plate & Cut 37.4 mg & Cut 71.3 mg & Cut 128.8 mg \\ \hline
$C$ & 0.5 & 0.47$\pm 0.05$ & 0.50$\pm 0.05$ & 0.47$\pm 0.04$ \\ \hline
$\Gamma _{I}$ & 12.6$\pm 0.3$ & 14.2$\pm 0.6$ & 17.6$\pm 0.7$ & 20.7$\pm 0.5$
\\ \hline
$\sigma _{I}$ & 2.8$\pm 0.3$ & 3.8$\pm 0.6$ & 5.0$\pm 0.7$ & 5.1$\pm 0.5$ \\ 
\hline
$\Gamma _{F}$ & 44.1$\pm 0.5$ & 48.1$\pm 0.8$ & 42.0$\pm 1.2$ & 40.1$\pm 0.7$
\\ \hline
$\sigma _{F}$ & 3.8$\pm 0.4$ & 5.3$\pm 0.7$ & 7.6$\pm 1.2$ & 7.2$\pm 0.7$ \\ 
\hline
\end{tabular}

\begin{figure}[tbp]
\end{figure}
FIG. 1. NNS distributions during the 2GOE-GOE crossover transition. The
histograms represent the numerical results of Leitner's \cite{leitner}
diagonalization of ensembles of two-block diagonal matrices, each of which
is a GOE, perturbed by random real matrices with strength parameters $%
\Lambda =4.05\times 10^{-4},8.11\times 10^{-3},4.05\times 10^{-2},8.11\times
10^{-1}$. The smooth curves are for the prediction of the proposed
independent-sequence model for symmetry breaking. The tuning parameter $%
f_{1} $ is calculated by using Eqs.(11) and (13) with the corresponding
values of interaction strength $\Lambda $.

{
\begin{figure}[tbp]
\end{figure}
}

FIG. 2. Resonance-width distributions for the acoustic resonances in intact
and cut three-leave clover-shaped plates, measured \ by Anderson et al. \cite
{andersen}, fitted by a sum of two Gaussian functions.

\begin{figure}[tbp]
\end{figure}
FIG. 3. NNS distributions for the acoustic resonances in intact and cut
three-leave clover-shaped plates measured \ by Anderson et al. \cite
{andersen}. The curves are the results of $\chi ^{2}$-fits by Eq. (14) for a
spectrum composed of two independent GOE sequences.

\begin{figure}[tbp]
\end{figure}
FIG. 4. Spectral rigidity $\Delta _{3}$ for the acoustic resonances in
intact and cut three-leave clover-shaped plates measured \ by Anderson et
al. \cite{andersen}. The curves are calculated for a superposition of two
independent GOE sequences with the same fractional densities that fit the
corresponding NNS distribution in Fig. 3.

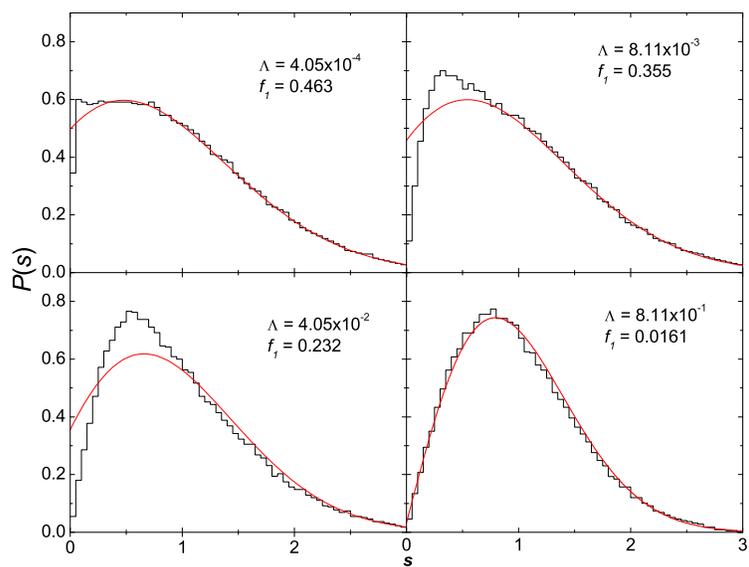
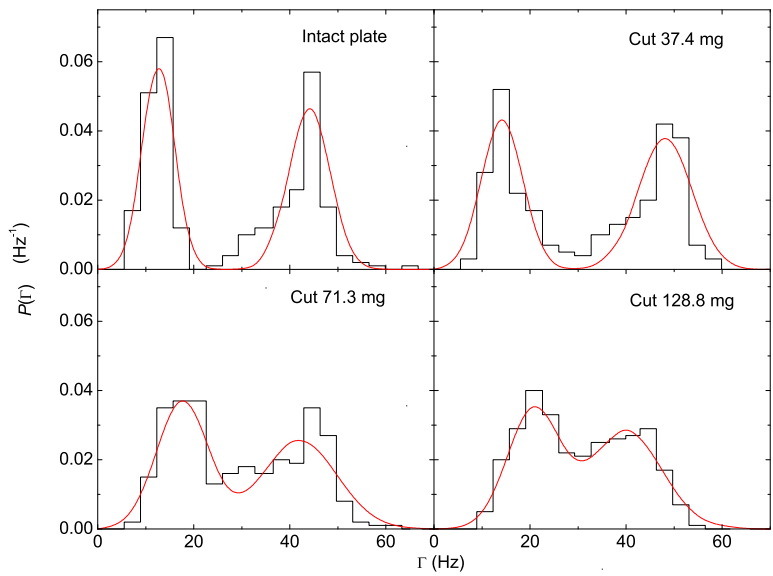
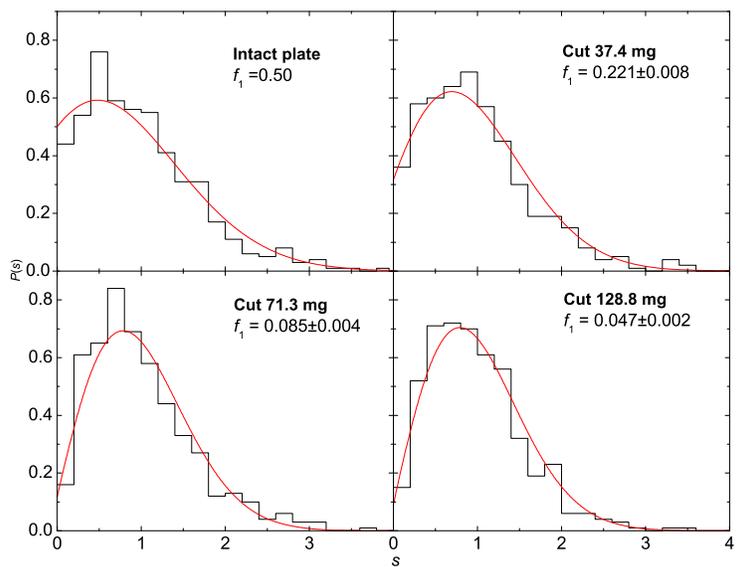
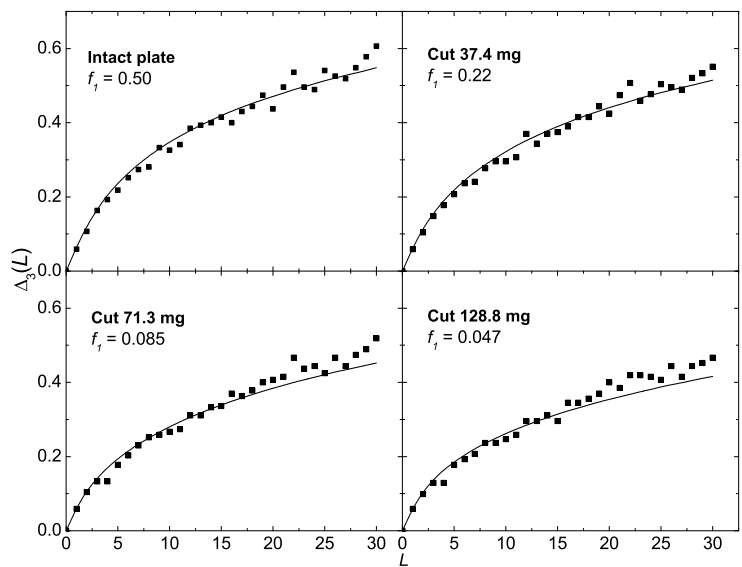
\end{document}